\documentclass[]{elsarticle}
\usepackage{lineno,hyperref}
\usepackage{amsmath}
\usepackage{bm}
\modulolinenumbers[5]

\journal{}

\begin{document}
\begin{frontmatter}

\title{Elastic stiffness tensors of Zr-$x$Nb alloy in presence of defects: A molecular dynamics study}

\author{Mohammad-Reza Basaadat}
\author{Mahmoud Payami\corref{mycorrespondingauthor}}

\address{School of Physics \& Accelerators, Nuclear Science and Technology Research Institute, AEOI, 
P.~O.~Box~14395-836, Tehran, Iran}

\cortext[mycorrespondingauthor]{Corresponding author}
\ead{mpayami@aeoi.org.ir}

\begin{abstract}
In a nuclear reactor, the Zr-$x$Nb alloy, which is used as a structural material in the core region, is irradiated by energetic particles that cause the atoms to be displaced from their lattice sites and giving rise to crystal defects.  
The local changes in the atomic arrangements lead to local deformations of the solid and thereby changes of its local mechanical properties. Understanding the mechanisms behind this evolution in the core region of a reactor, and its monitoring or controlling is a critical task in nuclear industry. 
In this work, using extensive molecular dynamics 
simulations, we have studied the effects of radiation damage on the local mechanical properties of Zr-$x$Nb alloy.    
In the first step, the effect of Nb-concentration on the mechanical stability of homogeneous 
Zr-$x$Nb alloy is investigated. In the second step, we have studied the local changes of the elastic constants due to local changes of the microstructure. These local changes include presence and accumulation of vacancies in the form of dislocation loops or voids, accumulation of Nb atoms in the form of clusters of different morphologies. 
This study covers both cases of $T=0^\circ$K and finite temperatures up to $T=600^\circ$K.
\end{abstract}

\begin{keyword}
Zr-$x$Nb alloy; Crystal defect; Elastic constant; Vacancy; Dislocation loop; Self-interstitial; Molecular dynamics 
\end{keyword}

\end{frontmatter}


\section{Introduction}\label{sec1}
Because of low cross-section for thermal-neutron capture, suitable corrosion-resistance in water, and good 
mechanical properties; zirconium and its alloys with niobium are widely used as structural materials in 
the core and fuel cladding of water-cooled nuclear 
reactors\cite{rickover1975history,lustman1955metallurgy,COX2005331}. 

Pure zirconium is realized in three different crystal structures: the $\alpha$-phase with an hcp lattice, 
being stable for temperatures lower than $860^\circ C$; and the $\beta$-phase, having bcc crystal structure, 
being 
stable at higher temperatures\cite{LEMAIGNAN2012217}. It is also observed that the $\omega$-phase, with 
hexagonal structure and space group $p6/mmm$, is obtained in a phase transition from $\alpha$-phase at 
pressure of 2-7 GPa\cite{xiahui}. 
On the other hand, pure niobium crystallizes with a bcc structure at normal conditions. 

The Zr-$x$Nb alloys at temperatures $T<800^\circ K$ are characterized by $\alpha$-phase crystalline 
structure\cite{fernandez1991thermodynamic}. 
Some of Zr-Nb based alloys that are widely used in nuclear reactors, are for example: Zr-$1\%$Nb-O, known 
as M5, used in fuel cladding materials\cite{mardon1997update}; Zr-2.5$\%$Nb-O alloys, used in CANDU 
reactors in Canada for pressure tube materials\cite{ALDRIDGE197232}; and a Zr-Nb-Sn-Fe alloy, known as 
ZIRLO, also used as a cladding material\cite{sabol1989zirconium}.

Some prior experimental and theoretical investigations on the mechanical properties of Zr-Nb 
alloys have been performed by other researchers. For example: the elastic moduli of Zr single crystal and their temperature 
dependence were experimentally studied early 
in 1964 by Fisher and coworker\cite{fisher1964}; the elastic constants of Nb single crystal were 
experimentally studied at 27$^\circ$~C by Bolef\cite{bolef61}; the elastic constants of Nb-rich alloys 
Nb-$x$Zr with bcc structure from 4.2$^\circ$~K up to room temperature were experimentally determined by 
Hayes and coworker\cite{hayes73} in 1973; and also from zero temperature to the melting point were studied 
both experimentally and theoretically by Ashkenazi and coworkers\cite{ashkenazi1978} in 1978. 
Varshni\cite{varshni1970}; 
Fast,~{\it et al.}\cite{fast1995}; and Olsson\cite{OLSSON2015361} have investigated the temperature 
dependence of 
elastic constants of zirconium; Peng,~{\it et al.} have studied 
the pressure effect on stabilities of self-interstitials in hcp-zirconium using first-principles  
method\cite{peng2014pressure}; Liu,~{\it et al.}\cite{liu2011first}; Wang,~{\it et 
al.}\cite{wangyi2017}; have calculated the elastic 
constants of niobium at normal pressure and high pressures, respectively; 
Kharchenko,~{\it et al.} have investigated\cite{kharchenko2012ab} the effect of Nb concentration on the 
lattice constants of Zr-$x$Nb;  
Wang,~{\it et al.}\cite{WANG201589}; Weck,~{\it et al.}\cite{weck2015}; Al-Zoubi,~{\it et 
al.}\cite{ALZOUBI2019273}; have investigated the elastic constants 
for
Zr-Nb alloys; Xin,~{\it et al.} have studied the point defect 
properties (such as formation energies) in hcp and bcc Zr with trace solute Nb by ab initio 
calculations\cite{XIN2009197}. However, to our knowledge, the mechanical properties of Zr-$x$Nb 
alloys in presence of defects have not been thoroughly investigated, which will be studied in the present work.  

In the core of a nuclear reactor, Zr-Nb alloys are irradiated by energetic neutrons, high-energy photons, 
or energetic ions.
When the colliding particles are fast enough, they can transfer sufficient kinetic energies to the nuclei 
of atoms of the alloy. Due to this energy transfer, the atoms are ejected from their crystal lattice 
sites, leaving behind vacancies (V), diffuse throughout the material, and will finally stop within the 
crystal as self-interstitials (SI)\cite{WOLFER20121}. The displaced atom with its corresponding 
vacancy forms a so-called Frenkel pair. 

At high enough temperatures and under the irradiation, the self-interstitial atoms and vacancies separately can diffuse throughout 
the crystal with different mobilities; 
giving rise to interstitial-vacancy recombinations or formation of larger atomic aggregates and vacancy 
clusters in the forms of dislocation loops or voids\cite{WOLFER20121,VARVENNE201465}. 
In addition, at intermediate temperatures the Nb atoms of the alloy may diffuse to a common region, 
leading to a local segregation of solute Nb atoms in a bcc structure\cite{ZINKLE201265,daria2017study}. 

The creation and evolution of the defects lead to local deformation of the alloy at those regions occupied by the 
defects. This deformation is due to local changes in the crystal lattice parameters or atomic arrangements 
which, in turn, lead to local changes in the elastic properties of the solid. 
Although the origins of such changes in the mechanical properties are known, the experimental 
determination of them in the core region of a reactor is a challenging task and therefore one must employ 
some theoretical models or resort to accurate simulation methods.

To determine the local changes in the mechanical properties of a defected crystal at different regions, we 
simulate that region in an MD box with a similar microstructure, and the size 
of the box is adopted sufficiently large so that the periodic images of the defects in simulation box do not 
interact.    

For large-scale molecular dynamics simulations, some interatomic potentials have already been developed 
and successfully used. For 
example: Mendelev,~{\it et. al.}\cite{mendelev2007development} using the embedded-atom method 
(EAM)\cite{DAW1993251} have developed 
Finnis-Sinclair type (``fs'') interatomic potential (Zr$\_$3.eam.fs) for pure $\alpha$-phase hcp 
zirconium, and 
have studied its phase transformations; Lin,~{\it et al.}\cite{Lin_2013} using the EAM 
have developed an n-body potential for a Zr-Nb system and calculated some ground-state 
properties; and very recently, Smirnova,~{\it et al.}\cite{SMIRNOVA2017259} by adding angular-dependent 
term to the EAM 
potential, 
have developed a new 
``adp'' interatomic potential (Zr$\_$Nb.adp.txt) and have shown that the structure and properties of all 
Nb and Zr phases existing in the Zr-Nb alloy were reproduced with good accuracy. This potential is 
extensively used in the present study.

In this work, using the ``adp'' interatomic potential within the LAMMPS molecular-dynamics (MD) code package\cite{lammps} we have performed extensive calculations on the structural and mechanical properties of  $\alpha$-phase Zr-$x$Nb systems with point and aggregated defects. 
The variations of the properties with respect to the temperature, up to working conditions of the reactor 
(600~$^\circ K$) were also studied. 

The structure of this paper is as follows. Section~\ref{sec2} is dedicated to the computational details; 
the calculation results are presented and discussed in section~\ref{sec3}; and in section~\ref{sec4}, we 
conclude this work. Finally, some useful relations employed for the calculations of mechanical properties are 
summarized in \ref{app}. 

\section{Computational Details}\label{sec2}
\begin{figure}
	\centering
	\includegraphics[width=\linewidth]{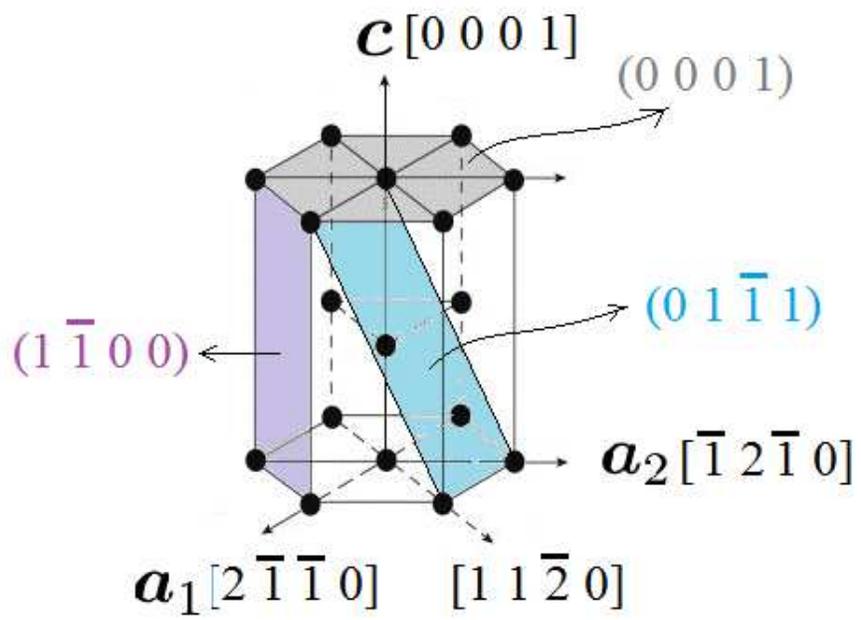}
	\caption{Crystal structure of Zr hcp lattice with some crystallographic planes and axes. }
	\label{fig1}
\end{figure}

To calculate the elastic constants of the solid with a given microstructure, 
we construct the simulation box with atomic configuration consistent with that microstructure and fully optimize the atomic positions as well as the cell geometry. 
In the next step, we take the relaxed supercell (simulation box) lattice vectors, denoted by $\left\{\boldsymbol{a}_1 , \boldsymbol{a}_2 , \boldsymbol{c} \right\}$, and apply deformations according to each of the 6 deformation gradients $\{\boldsymbol{G}_\alpha\}$ given by: 

\begin{equation}\nonumber
\boldsymbol{G}_1= \left[{\begin{matrix} 1 + \Delta & 0 & 0 \\ 0 & 1 & 0 \\ 0 & 0 & 1 \\ \end{matrix}}\right],
\boldsymbol{G}_2= \left[{\begin{matrix} 1 & 0 & 0 \\ 0 & 1 + \Delta  & 0 \\ 0 & 0 & 1 \\ \end{matrix}}\right],
\boldsymbol{G}_3= \left[{\begin{matrix} 1 & 0 & 0 \\ 0 & 1 & 0 \\ 0 & 0 & 1 + \Delta  \\ \end{matrix}}\right],
\end{equation}
\begin{equation}\label{deformgrad}
\boldsymbol{G}_4= \left[{\begin{matrix} 1 & \Delta & 0 \\ \Delta & 1 & 0 \\ 0 & 0 & 1 \\ \end{matrix}}\right],
\boldsymbol{G}_5= \left[{\begin{matrix} 1 & 0 & \Delta \\ 0 & 1 & 0 \\ \Delta & 0 & 1 \\ \end{matrix}}\right],
\boldsymbol{G}_6= \left[{\begin{matrix} 1 & 0 & 0 \\ 0 & 1 & \Delta  \\ 0 & \Delta & 1 \\ \end{matrix}}\right],
\end{equation}
and employ the Green-Lagrange strain tensor, $\boldsymbol{\varepsilon}^\alpha$ defined by:
\begin{equation}
\boldsymbol{\varepsilon}^\alpha=\frac{1}{2}(\boldsymbol{G}^T_\alpha \boldsymbol{G}_\alpha - \boldsymbol{I}).
\end{equation}
These 6 deformation gradients are applied one-by-one to the fully-relaxed structure such that only one independent deformation is considered at each time. For each of the 6 deformation modes, we perform positive and negative box displacements of "appropriate" magnitudes: $\Delta=-\delta, +\delta$. For each of deformed structures, the stress tensor, $\boldsymbol{\sigma}^\alpha$, is calculated by MD run, allowing for relaxation of all the atomic degrees of freedom. Modifying the indices of strain and stress tensors according to Voigt-notation: $11 \mapsto 1$, $22 \mapsto 2$, $33 \mapsto 3$, $23 \mapsto 4$, $13 \mapsto 5$, $12 \mapsto 6$, the components of elastic constants $C_{ij}^\alpha$ were determined from the linear equation:
\begin{equation}\label{sigmaceps}
\sigma_{i}^\alpha =\sum_{j=1}^6 C_{ij}^\alpha \epsilon_j^\alpha, \;\;\; (i,\alpha=1,2,\cdots ,6). 
\end{equation}
Finally, the nonzero components of elastic constants are determined from simple averaging over $C_{ij}(-\delta)$ and   
$C_{ij}(+\delta)$. In the hexagonal crystal system, it is common to take both lattice vectors $\boldsymbol{a}_1$ and $\boldsymbol{a}_2$ in the basal plane, whereas $\boldsymbol{c}$ orthogonal to the basal plane (See Fig.~\ref{fig1}) so that we have $C_{11} = C_{22}$. 

To run MD simulations at constant temperatures we utilize the Langevin thermostat\cite{schneider1978} with sufficiently small damping factor.
Before applying deformations to the system, we first find the equilibrium box shape and size by employing the Parrinello-Rahman method in NPT (isothermal-isobaric) statistical ensemble imposing the constraints on pressure $P=0$, and off-diagonal components $\sigma_{ij}=0$; and then using the resulting optimized parameters we apply the deformations and continue our simulations employing NVT ensemble to determine the resulting stress components. Afterwards, using Eq.(\ref{sigmaceps}), the components of elastic tensor are determined.  The equilibrations of such processes were achieved after 40~ps and 20~ps for NPT and NVT MD runs. The time-step was chosen as 0.001~ps.   

\begin{figure}
	\centering
	\includegraphics[width=\linewidth]{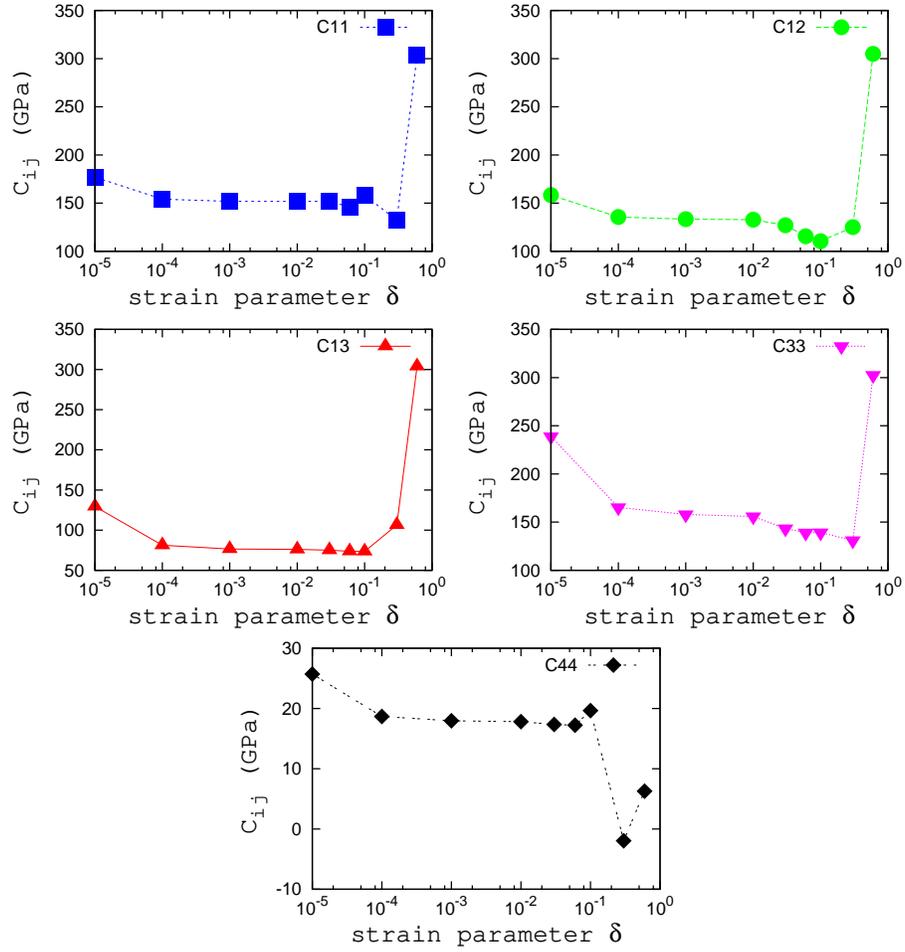}
	\caption{Calculated elastic constants as functions of strain parameter $\delta$ for Zr-1\%Nb system. The appropriate values of $\delta$, common to all elastic constants are in the range $10^{-4}\le\delta\le 10^{-2}$ which we chose to be $\delta=0.02$.}
	\label{fig2}
\end{figure}

To determine the "appropriate" magnitude of box displacements in each case, we have tried the values $\delta$=0.00001, 0.0001, 0.001, 0.01, 0.1; and plotted the resulting elastic constants $(C_{ij})$ as functions of $\delta$,  and determined the appropriate range of values for which the results are insensitive to the choice of displacement magnitude, and then chose a value with low computation costs. For example, in Fig.~\ref{fig2} we have shown the results for the case of Zr-1\%Nb system for which $\delta=0.02$ is a good choice.

Finally, the size of the simulation boxes were chosen to be sufficiently large to avoid the interactions between the specific defect in the box with those in the periodic image boxes. In addition, to simulate the properties at zero temperature, we have used a tiny value (near to zero) of $T=0.01$~K. So all listed properties at $T=0$ correspond to molecular dynamics simulations at this tiny value of temperature.

\section{Results and Discussions}\label{sec3}
To calculate the equilibrium geometry of pure $\alpha$-Zr (i.e., Zr-$x$Nb with $x=0\%$) at $T=0$, a simulation box comprising of an $8\times 8\times 6$ hcp unit cells containing 768 atoms was used. 
Then, allowing the changes of box size and shape as well as relaxing the atomic positions, we have fully optimized the geometry of the system and obtained the equilibrium lattice constants to be $a=3.22\AA$ and $c=5.17\AA$ with $c/a=1.60$, which are in good agreement with experimental values\cite{goldak1966lattice}: $a=3.23\AA$ and $c=5.15\AA$ with $c/a=1.59$.  In the second step, the elastic constants of Zr-xNb alloys with different Nb concentrations and morphologies are studied. In the next step, the effects of adding vacancies (V), self-interstitials (SI), vacancy dislocation loops, and voids on the elastic constants are investigated.    
\subsection{Homogeneous Nb distribution}
As was mentioned earlier, in the course of time the Nb concentration of Zr-$x$Nb may locally change and 
give rise to local changes 
of mechanical properties of the alloy. To see how the elastic constants depend on the local concentration $x$ of Nb 
atoms, we have calculated the elastic constants of homogeneously distributed Nb atoms in 
Zr-$x$Nb alloy for different values of $x=$0, 1, 2, 5, and $10\%$, where $x=0$ corresponds to pure Zr crystal. 

\begin{figure}
	\centering
	\includegraphics[width=\linewidth]{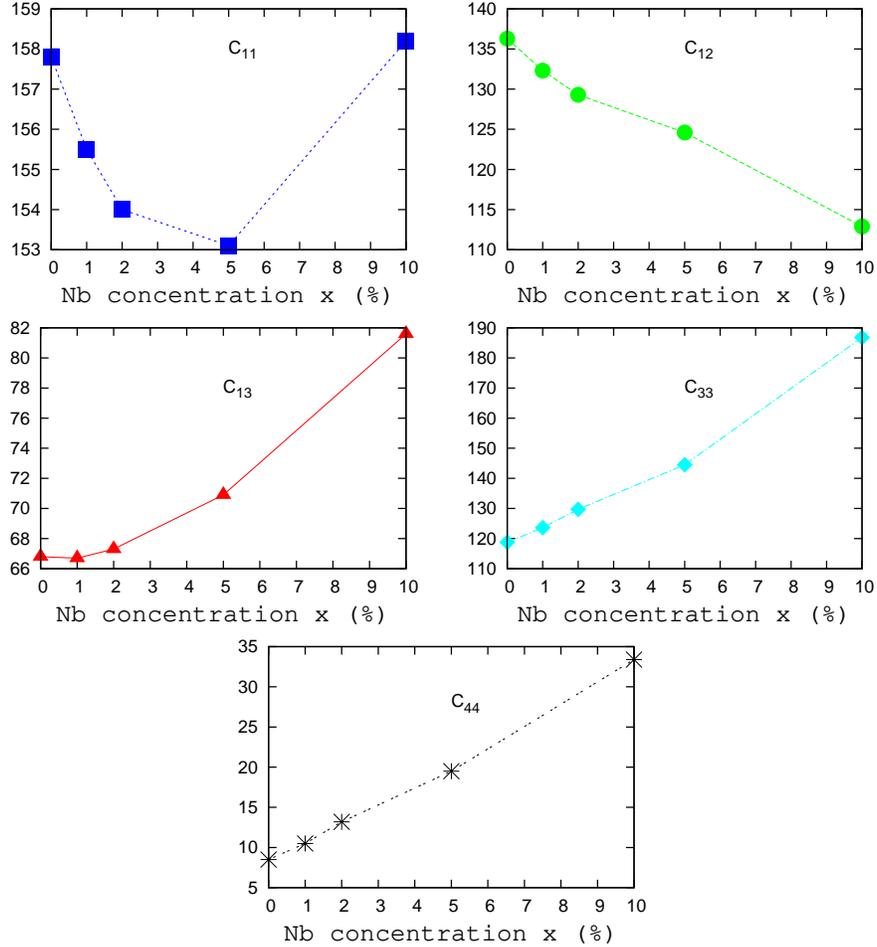}
	\caption{Elastic constants, in GPa, of homogeneous Zr-$x$Nb alloy for $x=0, 1, 2, 5, 10\%$ at $T=0$. The values corresponding to $x=0$ are for pure Zr crystal. }
	\label{fig3}
\end{figure}

The variations of $C_{ij}$'s with respect to Nb-concentration $x$ at $T=0$ are shown in Fig.~\ref{fig3}.
As is seen from Fig.~\ref{fig3}, the rate of changes (in $\%$) are 3, 17, 22, 57, and 325 for $C_{11}$, 
$C_{12}$, $C_{13}$, $C_{33}$, and $C_{44}$, respectively. Evidently, $C_{11}$ is the least sensitive 
elastic constant with $3\%$ of change 
for $0\le x\le 10$. The changes in other four elastic constants are relatively significant. $C_{12}$ shows 
decreasing behavior while other remaining three $C_{13}$, $C_{33}$, and $C_{44}$ increase with $x$. 

Using the conditions \ref{appeq16}, it was verified that Zr-$x$Nb alloys for $x=$1, 2, 5, and $10\%$ are 
elastically stable. We therefore conclude that in a Zr-$x$Nb alloy, there is no mechanical instability 
problems at inhomogeneous regions with $0\le x\le 10$. However, it should be mentioned that although the 
relatively high concentration of Nb at a region in Zr-$x$Nb alloy is mechanically harmless by itself, but 
this condition makes the region highly capable for hydrogen capture which decreases the elasticity and 
leads to development of cracks\cite{puls2012effect}. 

In Table~\ref{tab1}, we have listed the Young's, bulk, and shear moduli (See \ref{app}) as given by Voigt 
and Reuss theories for polycrystalline material. Inspecting the listed elastic moduli, we observe that 
they all increase with increasing the $x$ value in the alloy. This makes the alloy to become harder 
against any deformations. Increasing $x$ by $10\%$, Young's modulus increases by $115\%$ while the bulk and shear moduli increase by $12\%$ and $126\%$, respectively.
 
\begin{table}[]\small
\caption{Elastic moduli $E$, $K$, and $G$, in GPa, along with dimensionless Poisson's ratio $\nu$ and anisotropy index $A$. The subscripts ``V'' and ``R'' correspond to Voigt and Reuss theories, respectively. The superscripts ``G'' and ``U'' on $A$ correspond to empirical and universal measures, respectively.  }
 \centering
\resizebox{\columnwidth}{!} 
{
\begin{tabular}{lcccccccccccccc} \hline\hline
 Zr-$x$Nb   &$E_V$&$E_R$&&$K_V$&$K_R$&&$G_V$&$G_R$&&$\nu_V$&$\nu_R$&&$A^G$&$A^U$\\ \hline
 $x=0\%$ &44.6&31.9&&105.2&95.3&&15.6&11.0&&0.429&0.444&&0.171&2.169  \\ 
 $x=1\%$ &50.0&37.1&&107.3&99.5&&17.6&12.9&&0.422&0.438&&0.152&1.875  \\ 
 $x=2\%$ &55.2&42.5&&107.3&101.2&&19.5&14.8&&0.414&0.430&&0.136&1.628  \\
 $x=5\%$ &67.1&53.5&&109.3&106.2&&24.0&18.9&&0.398&0.416&&0.119&1.378  \\
 $x=10\%$&95.9&82.8&&117.3&117.2&&35.2&29.9&&0.364&0.382&&0.081&0.876  \\ \hline\hline
\end{tabular}
}
	\label{tab1}
\end{table}

Looking at the Poisson's ratio $\nu$ and anisotropy index $A$ values listed in Table~\ref{tab1}, we notice 
that both of them decrease by increasing the Nb concentration $x$ in the alloy. That is, by $10\%$ 
increase of $x$, the Poisson's ratio decreases by $15\%$ from 0.429 to the value 0.364, while the ``anisotropy 
index'' decreases by $60\%$ from 2.169 to the value 0.876.    

\begin{figure}
	\centering
	\includegraphics[width=\linewidth]{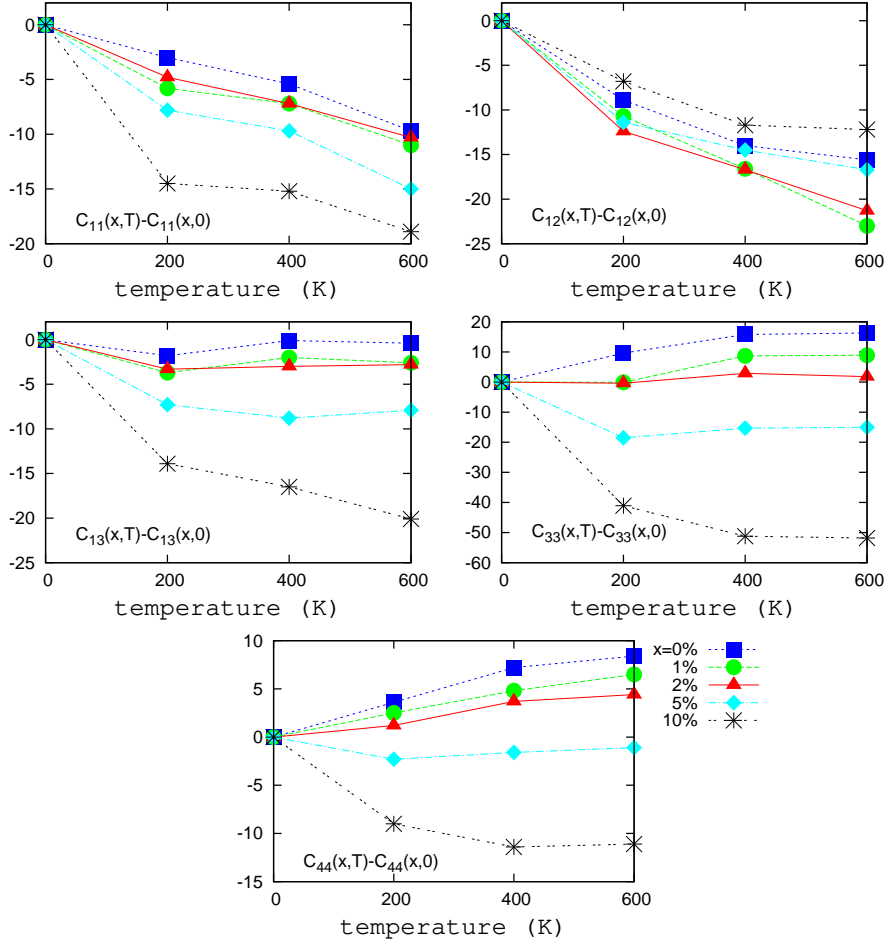}
	\caption{Changes of elastic constants, $C_{ij}(x,T)-C_{ij}(x,0)$, in GPa, of homogeneous Zr-$x$Nb alloy for $x=0, 1, 2, 5, 10\%$ as 
	functions of temperature. $C_{ij}(x,0)$ corresponds to the values at $T=0$ (See Fig.~\ref{fig3}). }
	\label{fig4}
\end{figure}

Because of thermal expansion of the alloy and thermal vibrations of atoms, the $C_{ij}$'s undergo changes 
with 
increasing the temperature. They generally decrease with temperature, because the thermal expansion 
decreases the interacting forces between atoms. In Fig.~\ref{fig4} we have shown the 
changes with temperature of elastic constants for 
homogeneous Zr-$x$Nb including $x$=0, 1, 2, 5, and 10$\%$. These changes are measured from the corresponding values at $T=0$.    
From Fig.~\ref{fig4} we notice some general features that are worth mentioning: {\it i})- The elastic 
constants $C_{11}$ and $C_{12}$ have similar behaviors in that for all studied concentrations $x$, they show decreasing behaviors with temperature, while for other elastic constants, the decreasing behavior is remained only for higher concentrations of $x=5,\;10\%$; {\it ii})- $C_{13}$, 
$C_{33}$, and $C_{44}$ increase with temperature or remain almost constant for lower Nb-concentrations $x=0,\;1,\;2\%$. In other words, the increasing behaviors of $C_{13}$, $C_{33}$, and $C_{44}$ for lower $x$ values changes to decreasing behaviors for higher Nb concentrations.
A more careful inspection of Fig.~\ref{fig4} reveals that except for $C_{12}$, other elastic constants 
have highest variations in the interval $0<T<600^\circ K$ for $x=10\%$. In other words, the highest 
changes correspond to $x=10\%$ for $C_{11}$, $C_{13}$, $C_{33}$, $C_{44}$ by respective values of $11\%$, 
$24\%$, $28\%$, 
$32\%$.  

\subsection{Inhomogeneous Nb distribution}
\begin{figure}
	\centering
	\includegraphics[width=\linewidth]{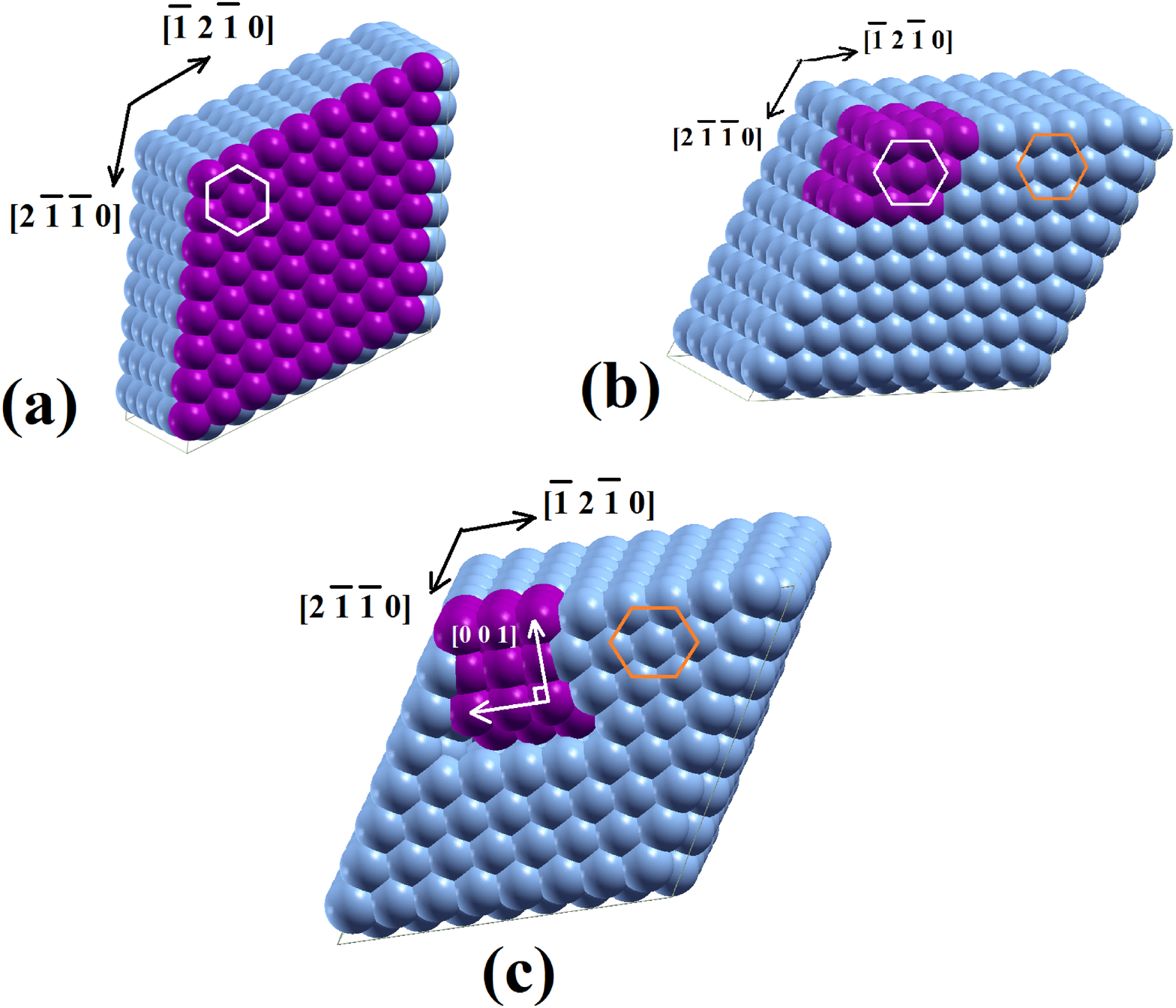}
	\caption{Different local Nb-inhomogeneities in Zr-$x$Nb alloy: (a)- planar Nb-aggregate, (b)- 
	three-dimensional hcp Nb-aggregate, and (c)- three-dimensional bcc Nb-aggregate. The average local Nb concentration is $x=8\%$. Zr and Nb atoms are specified with dark blue and violet colored balls, respectively. }
	\label{fig5}
\end{figure}

The local inhomogeneity of Zr-$x$Nb alloy, produced by Nb aggregates, may be realized in different 
morphologies. To show how the morphologies can affect the local mechanical properties, we have considered 
three examples of different local inhomogeneities: planar Nb-aggregate (parallel to basal plane), three-dimensional hcp Nb-aggregate, and three-dimensional bcc Nb-aggregate (See Fig.~\ref{fig5}).  
As shown in Fig.~\ref{fig5}(a), to create a planar Nb-aggregate, we have replaced one layer (in 
basal plane) of 
Zr atoms in the simulation box with one layer of Nb atoms.
To create a 54-atom hcp-aggregate, as in Fig.~\ref{fig5}(b) we have just replaced the Zr atoms in a $3\times 3\times 3$ supercell with Nb atoms.  However, in Fig.~\ref{fig5}(c) the replaced Nb-aggregate has bcc structure as is realized in precipitates.
The elastic constants obtained from the MD runs at $T=0$ for these three examples of Nb-inhomogeneities 
are presented in Table~\ref{tab2}.  
\begin{table}[]\small
\caption{Elastic constants in GPa (at $T=0$) of Zr-$x$Nb alloy of local Nb-inhomogeneity for different 
morphologies of Nb aggregates: planar, three-dimensional hcp, and three-dimensional bcc. The Nb average 
concentration in the simulation box is $x=8\%$.  }
 \centering
{
\begin{tabular}{lccccc} \hline\hline
  shape  & $C_{11}$ & $C_{12}$ & $C_{13}$ & $C_{33}$ & $C_{44}$ \\ \hline
 planar  & 157.5 & 121.4 & 67.8 & 134.0 & 9.3 \\ 
 hcp     & 151.8 & 114.6 & 65.5 & 136.1 & 14.9 \\
 bcc     & 143.8 & 116.6 & 69.2 & 116.3 & 9.8 \\ \hline\hline
\end{tabular}
}
	\label{tab2}
\end{table}
In Table~\ref{tab2}, the results show that $C_{11}$ decrease as we go from planar to hcp to bcc, whereas $C_{12}$ and $C_{13}$ decrease from planar to hcp but increase from hcp to bcc. Conversely, for $C_{33}$ and $C_{44}$ when we go from planar to hcp, they increase while by going from hcp to bcc they show decreasing behaviors. Quantitatively, the morphology of the Nb aggregates have contribution in the changes of elastic constants by at most $\sim$15GPa which corresponds to $C_{11}$ and $C_{33}$ whereas the least sensitive one is $C_{13}$ which varies at most by $\sim$2GPa.

\begin{figure}
	\centering
	\includegraphics[width=\linewidth]{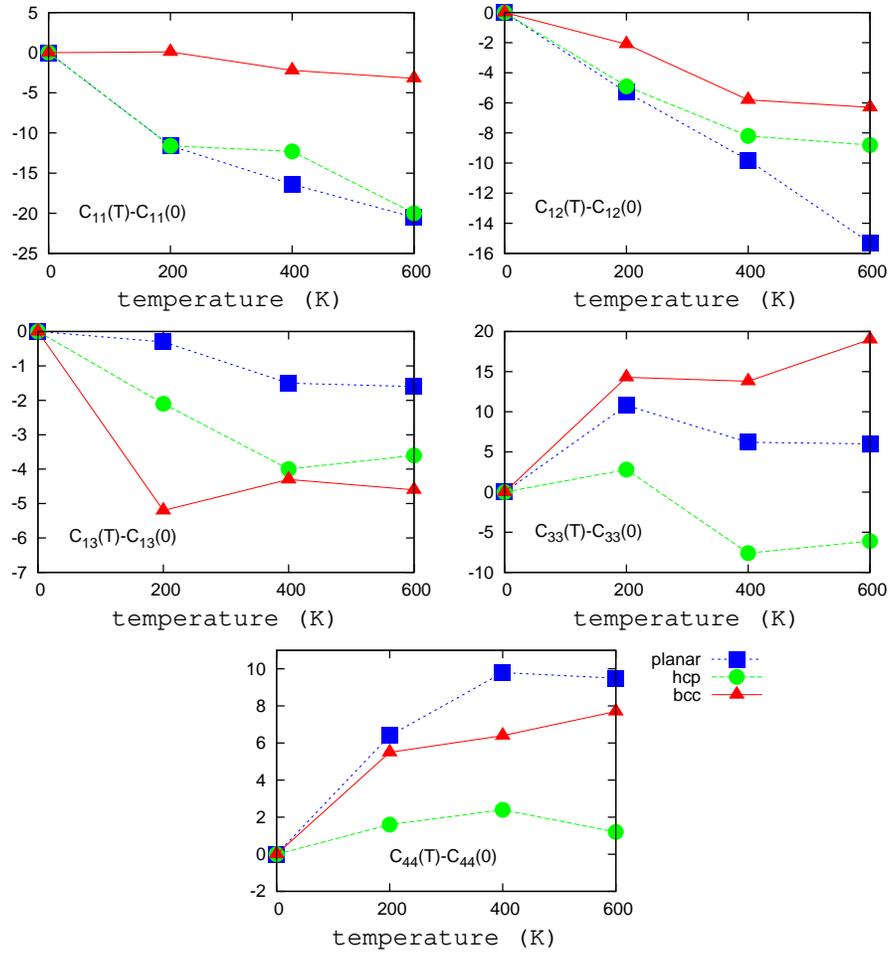}
	\caption{Temperature dependence of elastic constants for different morphologies of Nb inhomogeneity. 
	Planar, hcp, and bcc correspond to different Nb-inhomogeneity morphologies. See 
	Fig.~\ref{fig5}.  }
	\label{fig6}
\end{figure}

In Fig.~\ref{fig6}, we have plotted the temperature dependence of elastic constants for the systems shown in Fig.~\ref{fig5}. The reference points $C_{ij}(0)$ were taken as the corresponding values at $T=0$ (as tabulated in Table~\ref{tab2}). The variations with respect to temperature are at most $\sim$20GPa which for $C_{11}$ belongs to planar and hcp inhomogeneities while for $C_{33}$ it belongs to local bcc Nb-precipitate.     

\subsection{Vacancies, self-interstitials, vacancy dislocation loops, and voids}
\begin{figure}
	\centering
	\includegraphics[width=\linewidth]{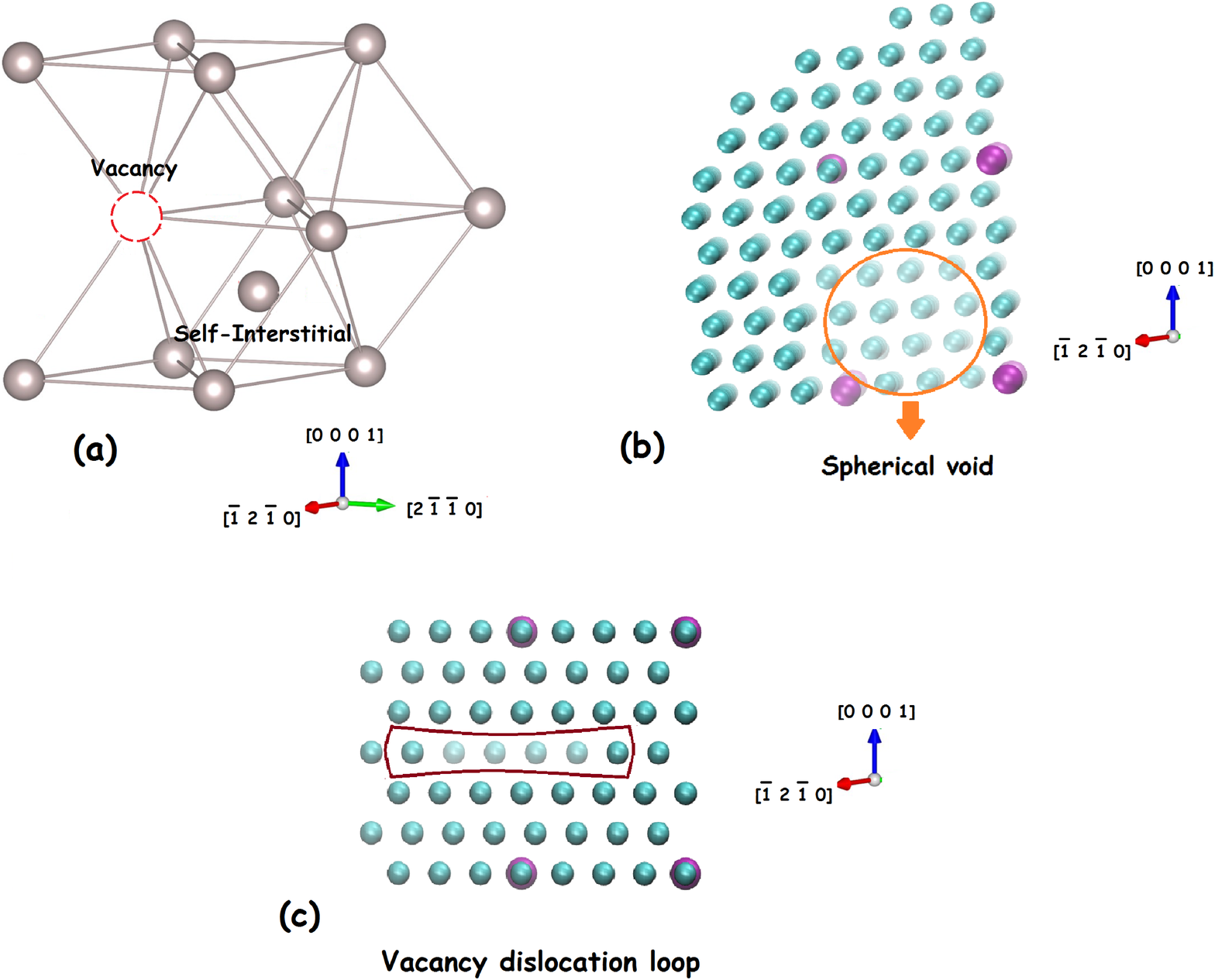}
	\caption{Crystal defects as: (a)-single-atom vacancy and self-interstitial, (b)- void, that forms from joining the single-atom vacancies, and (c)- vacancy dislocation loop which is the most probable form of vacancy-aggregate. }
	\label{fig7}
\end{figure}

Energetic particles when colliding the nuclei of atoms in a solid, are capable to eject atoms from their 
crystal lattice sites, leaving behind vacancies. On the other hand, the ejected atom move throughout the crystal and may stop somewhere as an interstitial atom [See Fig.~\ref{fig7}(a)].    
The vacancies can diffuse throughout the crystal to join and give rise to formation of large vacancy 
clusters, also called voids [(See Fig.~\ref{fig7}(b)] or in the form of vacancy dislocation loops as shown in Fig.~\ref{fig7}(c). 

The existing internal stress fields of some extended defects may cause the single-atom vacancies and SI's or their clusters move in such a way that lead to partial segregation of atoms. However, in our study we do not consider their migration but have first investigated how the homogeneous distribution of V's or SI's can affect the local values of elastic constants.

\begin{figure}
	\centering
	\includegraphics[width=\linewidth]{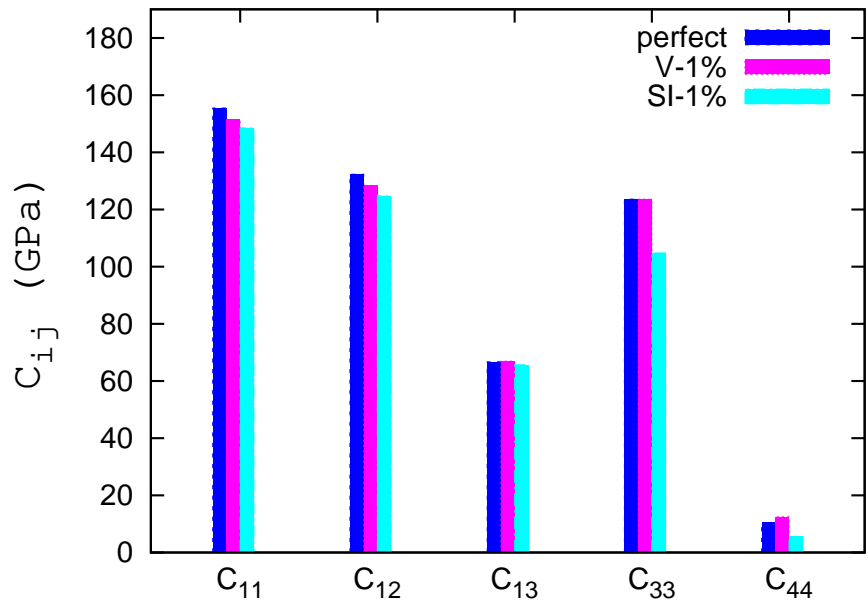}
	\caption{Elastic constants of two Zr-$1\%$Nb systems, one with 1$\%$ homogeneously-distributed V's and 
	other with 1$\%$ homogeneously-distributed SI's. The violet, blue, and cyan bars correspond to Zr-$1\%$Nb system with $1\%$V's, perfect, and Zr-$1\%$Nb system with $1\%$ SI's, respectively. }
	\label{fig8}
\end{figure}

To simulate the Zr-$1\%$Nb alloy with $1\%$-V, we have removed homogeneously 8 Zr-atoms from the supercell containing 768 atoms, and for the $1\%$-SI, we have added 8 Zr-atoms homogeneously in the octahedral sites, so that the former and latter supercells have 768 and 776 atoms, respectively. Then the systems were geometrically optimized and the corresponding measured lattice constants are listed in Table~\ref{tab3}.
\begin{table}[]\small
	\caption{Average lattice constants $a$ and $c$, in \AA, as well as $c/a$ for pure Zr, perfect Zr-$1\%$Nb, Zr-$1\%$Nb + $1\%$-V, and Zr-$1\%$Nb + $1\%$-SI systems at $T=0$, respectively.}
	\centering
	{
		\begin{tabular}{lcccc} \hline\hline
			lattice parameter  & Zr-$0\%$Nb & Zr-$1\%$Nb & Zr-$1\%$Nb + $1\%$-V & Zr-$1\%$Nb + $1\%$-SI \\ \hline
			$a$                & 3.215      & 3.213      & 3.210                & 3.215                 \\ 
			$c$                & 5.173      & 5.164      & 5.156                & 5.221                 \\
			$c/a$              & 1.609      & 1.607      & 1.606                & 1.624              \\ \hline\hline
		\end{tabular}
	}
	\label{tab3}
\end{table}
From the results listed in Table~\ref{tab3}, we see that adding $1\%$-V to the Zr-$1\%$Nb system reduces (on the average) the values of $a$ and $c$ by 0.003 and 0.008\AA, respectively; while adding $1\%$-SI increases them by 0.002 and 0.057\AA, respectively. From these results we expect the interaction between atoms in the SI case to reduce compared to the other two cases, which in turn reduces the corresponding elastic constants. To verify this fact, we have obtained and compared the elastic constants for the three cases in Fig.~\ref{fig8}.   
As is seen from Fig.~\ref{fig8}, by adding the point defects, all $C_{ij}$'s except for $C_{44}$ were decreased or not changed with respect to the perfect case, while $C_{44}$ increases in the case of V defect. In addition, as expected before, adding SI point defects to the system lead to higher changes in the $C_{ij}$ values. 

Finally, as was mentioned, the vacancy point defects may migrate under the influence of stress fields of extended defects and join together to form vacancy clusters in the forms of dislocation loops or voids. To understand how these defects may affect the local mechanical properties, for a given concentration of vacancies, we have considered four arrangements of vacancy defects for a Zr-$1\%$Nb system. In the first case, we have considered the single-atom vacancies (with $1\%$ concentration) that are homogeneously distributed throughout the Zr-$1\%$Nb alloy; in the second case, the vacancies that 
are joined to give rise to a maximum-sized dislocation-loop in basal plane; in the third case, dislocation-loop in prismatic plane; and in the fourth case, in the form of void. The corresponding elastic constants are listed in Table~\ref{tab4}. 
\begin{table}[]\small
	\caption{Elastic constants in GPa (at $T=0$) of Zr-$1\%$Nb alloy containing 1$\%$-V in different clustering forms: homogeneous, vacancy cluster in basal plane ($\langle c \rangle$-loop), vacancy cluster in prismatic plane ($\langle a \rangle$-loop), and vacancy void.}
	\centering
	{
		\begin{tabular}{lccccc} \hline\hline
		vacancy type	& $C_{11}$ & $C_{12}$ & $C_{13}$ & $C_{33}$ & $C_{44}$ \\ \hline
		no vacancy      & 155.5 & 132.3 & 66.7 & 123.6 & 10.5 \\ 
		homogeneous     & 151.5 & 128.0 & 66.3 & 122.4 & 12.2 \\ 
		basal plane     & 151.7 & 126.8 & 64.4 & 115.6 & 11.1 \\
		prismatic plane & 149.9 & 124.5 & 65.2 & 121.1 & 11.2 \\
		void            & 150.5 & 126.3 & 64.6 & 118.4 & 10.8 \\ \hline\hline
		\end{tabular}
	}
	\label{tab4}
\end{table}
From Table~\ref{tab4}, two points are worth mentioning: First, introduction of $1\%$-V into Zr-$1\%$Nb changes the values of $C_{11}$, $C_{12}$, $C_{13}$, $C_{33}$, and $C_{44}$ at most by 4, 6, 3, 6, and 16$\%$, respectively; Second, as expected, vacancy cluster in the form of $\langle c \rangle$-loop decreases the $C_{33}$ with respect to perfect alloy, while in the case of $\langle a \rangle$-loop vacancy, the $C_{11}$ is decreased. 

\subsection{Eisentropic elastic constants}
In this study, we have calculated the ``isothermal'' elastic constants at finite temperatures. However, in experimental data usually the ``isentropic'' values are reported and therefore it is necessary to calculate the corresponding conversion values. For hcp materials, an approximate relation between these two quantities are given by\cite{DAVIES19741513}:

\begin{equation}\label{aa}
C_{ij}^S(T)=C_{ij}^T(T)+\Delta_{ij},
\end{equation}
where $C_{ij}^S(T)$ and $C_{ij}^T(T)$ are ``eisentropic'' and ``isothermal'' elastic constants, respectively, and 
\begin{equation}\label{bb}
\Delta_{ij}={T\lambda_i\lambda_j\over \rho c_\epsilon}
\end{equation}
with 
\begin{equation}\label{cc}
\lambda_1=\lambda_2=\alpha_a\left(C_{11}^T+C_{12}^T\right)+\alpha_cC_{13}^T\textnormal{  ,  } 
\lambda_3=2\alpha_aC_{13}^T+\alpha_cC_{33}^T.
\end{equation}

In Eq.(\ref{cc}), $\alpha_a$ and $\alpha_c$ are the linear thermal expansion coefficients in the 
directions of ``$a$'' and ``$c$'' axes of an hcp crystal, respectively. $\rho$ and $c_\epsilon$ are the 
density and the specific heat at constant strain, respectively. In the $\alpha$-phase Zr-$x$Nb alloy we 
assume that the corresponding symmetries for an hcp perfect crystal are still valid and using the above 
relations we estimate the conversion values. The experimental value of $c_\epsilon$ for pure zirconium is 
reported\cite{guillermet1987critical} to be arround 270~Jkg$^{-1}$K$^{-1}$, and the linear thermal expansion 
coefficients $\alpha_a$ and $\alpha_c$ were experimentally reported\cite{skinner1953} to be of order 
$10^{-6}$~K$^{-1}$. From our calculations, in the range of $T=0^\circ$K and $T=600^\circ$K, the average specific heat $\bar{c}_\epsilon$ of pure $\alpha$-Zr at $P=0$ is obtained to be 273~Jkg$^{-1}$K$^{-1}$; and for the linear expansion coefficients at the same temperature range, we have obtained the average values of $\bar{\alpha}_a=2.3\times 10^{-6}$K$^{-1}$ and $\bar{\alpha}_c=7.0\times 10^{-6}$K$^{-1}$. Putting these values in the 
conversion formulas, we estimate 
the value of order less than $\sim 1$GPa which are practically insignificant. In fact, the conversion 
values would be significant for materials with high elastic constants, high thermal expansion 
coefficients, low densities, and low specific heats. 

\section{Conclusions}\label{sec4}
In this study, employing an extensive MD simulations, the effects of different crystal defects on the elastic stiffness coefficients of Zr-$1\%$Nb alloy were investigated.      
The defects considered in this study include: the local changes of Nb concentration, Nb clusters with different morphologies, single-atom vacancies and self-interstitials, vacancy clusters in the form of dislocation loops and voids. 
The local changes of Nb concentration up to $10\%$ showed that $C_{11}$, $C_{12}$, $C_{13}$, $C_{33}$, and $C_{44}$
undergo changes of at most 3, 17, 22, 57, and 325$\%$, respectively and that $C_{11}$ was the least sensitive 
elastic constant. In addition, it was shown that Zr-$x$Nb alloys for $x=$1, 2, 5, $10\%$ were 
elastically stable, and on the other hand, increasing the $x$ value in the alloy makes it become harder 
against any deformations.
Moreover, studying the temperature effects on Zr-$x$Nb showed that in the interval $0<T<600^\circ K$, except for $C_{12}$, the highest 
changes in elastic constants correspond to $x=10\%$ for $C_{11}$, $C_{13}$, $C_{33}$, $C_{44}$ by respective values of $11\%$, $24\%$, $28\%$, $32\%$.  
In the context of Nb aggregates, it was shown that at $T=0^\circ$K, the morphology has contributions by at most $\sim$15GPa in the elastic constants, which corresponds to $C_{11}$ and $C_{33}$; while, the variations with respect to temperature were at most by $\sim$20GPa for $C_{11}$ in planar form as well as hcp Nb-cluster, and for $C_{33}$ in the bcc Nb-precipitate. 
As to single-atom vacancies and self-interstitials, we have shown that adding $1\%$-V to the Zr-$1\%$Nb system, the values of $a$ and $c$ reduces by 0.003 and 0.008\AA, respectively; while addition of $1\%$-SI increases them by 0.002 and 0.057\AA, respectively. 
To understand the effects of vacancy clusters in the forms of dislocation loops or voids on the local mechanical properties, we had considered four arrangements of vacancy defects for a Zr-$1\%$Nb system with $1\%$-V. In the first case, the homogeneously-distributed single-atom vacancies were considered; in the second case, the maximum-sized dislocation-loop in basal plane; in the third case, dislocation-loop in prismatic plane; and in the fourth case, the maximum-sized void were considered. The results showed that firstly, introducing of $1\%$-V into Zr-$1\%$Nb changes the values of $C_{11}$, $C_{12}$, $C_{13}$, $C_{33}$, and $C_{44}$ at most by 4, 6, 3, 6, and 16$\%$, respectively; secondly, the vacancy cluster in the form of $\langle c \rangle$-loop decreases the $C_{33}$ with respect to perfect alloy, while in the case of $\langle a \rangle$-loop vacancy, the value of $C_{11}$ is decreased. 
Finally, it was shown that the difference between the calculated isothermal elastic constants and the experimental isentropic values were of order less than $\sim 1$GPa for Zr-$x$Nb alloy, which were practically insignificant and can be ignored. 

\section*{Acknowledgement} 
This work is part of research program in School of Physics and Accelerators, NSTRI, AEOI.  

\appendix
\section{Definition of quantities used in mechanical properties}\label{app}
\subsection{Elastic constants}
Deforming a crystal with a given strain $\epsilon_i$, some forces appear tending to bring it to the equilibrium configuration (Voigt notation is used)\cite{nye1985physical,wallace1998thermodynamics,mouhat14}. Stress tensor is used for the description of these forces. For small values of strain, the stress tensor becomes proportional to the strain so that we will have:

\begin{equation}\label{appeq1}
\sigma_{i} =\sum_{j=1}^6 C_{ij} \epsilon_j, \;\;\; (i=1,2,\cdots ,6) 
\end{equation}
The inverse of the Eq.(\ref{appeq1}) is given by:
\begin{equation}\label{appeq2}
\epsilon_{i} =\sum_{j=1}^6 S_{ij} \sigma_j, \;\;\; (i=1,2,\cdots ,6) 
\end{equation}

\noindent Here, $C_{ij}$ and $S_{ij}$ define the elastic and the compliance tensors . At zero temperature, we can write:

\begin{equation}\label{appeq3}
C_{ij}= {\partial \sigma_i \over \partial \epsilon_{j}},\;\;\;\;S_{ij}= {\partial \epsilon_i \over \partial \sigma_{j}}
\end{equation}
The nonzero components of the elastic constant tensor can be derived using group theory. 
For hexagonal lattice, we have five nonzero independent components:
\begin{equation}\label{appeq5}
C_{11}=C_{22},\;\;\;C_{12},\;\;\;C_{13}=C_{23},\;\;\;C_{33},\;\;\;C_{44}=C_{55},\;\;\;C_{66}={1 \over 2}(C_{11}-C_{12}).
\end{equation}

\subsection{Elastic moduli}\label{elasticmoduli}
Young's modulus $E$ describes the tensile elasticity, i.e., the tendency of an object to deform along an 
axis when opposing forces are applied along that axis; and it is defined as the ratio of tensile stress to 
tensile strain. 
The shear modulus (also called modulus of rigidity) $G$ describes an object's tendency to shear (i.e., 
undergo shape deformation at constant volume) when applied upon opposing forces; and it is defined as the 
ratio of shear stress to shear strain\footnote{Care must be taken not to confuse the symbol $G$ used here with those $\boldsymbol{G}_\alpha$ used for deformation gradients.}. 
The bulk modulus, denoted by $K$, describes volumetric elasticity, i.e., the tendency of a solid to deform 
in all directions when it is uniformly loaded in all directions; and is defined as the ratio of volumetric 
stress to volumetric strain. Bulk modulus is the inverse of compressibility and is the extension of 
Young's modulus to three dimensional case.

The bulk and rigidity moduli for a polycrystalline material, that is macroscopically isotropic, are given 
from Voigt and Reuss theories by\cite{hill1952}:
\begin{equation}\label{appeq6}
K_V = {1\over 9}(C_{11}+C_{22}+C_{33})+{2\over 9}(C_{12}+C_{23}+C_{13})
\end{equation}
\begin{equation}\label{appeq7}
G_V = {1\over 15}(C_{11}+C_{22}+C_{33})-{1\over 15}(C_{12}+C_{23}+C_{13})+{1\over 5}(C_{44}+C_{55}+C_{66}). 
\end{equation}
\begin{equation}\label{appeq8}
{1 \over K_R} = (S_{11}+S_{22}+S_{33})+2(S_{12}+S_{23}+S_{13})
\end{equation}
\begin{equation}\label{appeq9}
{1 \over G_R} = {4 \over 15}(S_{11}+S_{22}+S_{33})-{4 \over 15}(S_{12}+S_{23}+S_{13})+{3 \over 15}(S_{44}+S_{55}+S_{66}).
\end{equation}

For any stress and strain, we have:
\begin{equation}\label{appeq10}
K_R\le K\le K_V; \;\;\;\;G_R\le G\le G_V.
\end{equation}

Poisson's ratio $\nu$ is defined as the ratio of transverse contraction strain to longitudinal extension strain in the direction of a stretching force. The Poisson's ratio $\nu$ and Young's modulus $E$ for each scheme are given by:
\begin{equation}\label{appeq11}
\nu={1\over 2}\left[1-\frac{3G}{3K+G}\right] \;\;\;\;\;\frac{1}{E}=\frac{1}{3G}+\frac{1}{9K}.
\end{equation}
From Eq.(\ref{appeq10}) it is clear that:
\begin{equation}\label{appeq12}
E_R\le E\le E_V.
\end{equation}

\subsection{Anisotropy index}
The Chung-Buessem empirical measure for anisotropy of shear modulus is given by\cite{chung1967}:
\begin{equation}\label{appeq13}
A^G={G_V-G_R\over G_V+G_R},
\end{equation}
and the universal index of anisotropy is defined as\cite{ranganathan14}:
\begin{equation}\label{appeq14}
A^U=5({G_V \over G_R})+({K_V\over K_R})-6.
\end{equation}

\subsection{Elastic stability conditions}
The necessary and sufficient elastic stability conditions for hexagonal crystal are given by\cite{mouhat14}:
\begin{eqnarray}\label{appeq16}
{\rm Hexagonal:}\;\;\;\;\;C_{11}>  \mid C_{12} \mid,\;\;\;\;(C_{11}+C_{12})C_{33}> 2C_{13}^2,\;\;\;\\ \nonumber C_{44}>0,\;\;\;\;   C_{66}>0.\;\;\;\;\;\;\;\;\;\;\;\;\;\;\;\;\; 
\end{eqnarray}

\section*{Declaration of competing interests}
The authors declare no affiliations with or involvement in any organization or entity with any financial or non-financial interest in the subject matter or materials discussed in this research paper.

\section*{Data availability }
The raw or processed data required to reproduce these results can be shared upon an email to corresponding author (mpayami@aeoi.org.ir).
 
\bibliography{basaadatpayami-98.04.29}

\end{document}